\documentclass[aps,prd,twocolumn,showpacs,groupedaddress]{revtex4}
\usepackage{graphicx}
\usepackage{epsfig}
\usepackage{bm}
\usepackage{amssymb}

\newcommand{\lag}{\mathcal L}
\newcommand{\be}{\begin{eqnarray}}
\newcommand{\ee}{\end{eqnarray}}

\begin{document}


\title{QHD equation of state for strongly magnetized neutron stars}

\author{Chung-Yeol Ryu}
\affiliation{Department of Physics,
Soongsil University, Seoul 156-743, Korea}

\author{Myung-Ki Cheoun}
\thanks{cheoun@ssu.ac.kr}
\affiliation{Department of Physics,
Soongsil University, Seoul 156-743, Korea}

\author{ Toshitaka Kajino}
\affiliation{National Astronomical Observatory of Japan, 2-21-1 Osawa,
Mitaka, Tokyo 181-8588, Japan \\
Department of Astronomy, Graduate School of Science, University of
Tokyo, Hongo 7-3-1, Bunkyo-ku, Tokyo 113-0033, Japan }

\author{Tomoyuki~Maruyama}
\affiliation{College of Bioresource Sciences, Nihon University,
Fujisawa 252-8510, Japan}


\author{Grant J. Mathews}
\affiliation{Center for Astrophysics, Department of Physics, University of Notre Dame,
IN 46556, USA}

\begin{abstract}
We investigate the quantum hadrodynamic equation of state for  neutron stars (with and without including hyperons)
in the presence  of strong magnetic fields. The deduced masses and radii  are consistent  with   recent observations of high mass neutron stars even in the case of hyperonic nuclei for sufficiently strong magnetic fields. The
calculated adiabatic index and the moments of inertia for  magnetized neutron stars  exhibit rapid changes with density.  This may  provide some
insight into the mechanism of  star-quakes and flares  in  magnetars.
\end{abstract}


\maketitle

\section{Introduction}
Soft $\gamma$-ray repeaters (SGRs)
and anomalous $X$-ray pulsars (AXPs) are believed to be evidence for magnetars, i.e. neutron
stars with surface
magnetic fields of $10^{14} \sim 10^{15}$ G
\cite{cardall2001}. In the interior of these magnetic  neutron stars, the magnetic
field strength could be as much as  $10^{18}$ G according to the scalar
virial theorem. Such strong magnetic fields may affect the properties
of neutron stars such as the relative populations of various particles, the equation of
state (EOS), and the mass-radius relation. Many studies of dense nuclear
 matter in the presence of  strong magnetic fields have been reported \cite{band1997,brod2000,su2001,dey2002,shen2006,shen2009,panda2009,Ryu:2010zzb}. These
works have considered the electromagnetic interaction,
the Landau quantization of charged particles, and the baryon anomalous
magnetic moments (AMMs).
However, a detailed analysis of  neutron stars with strong magnetic
fields is still an area of active research.

Recently \cite{palmer2005}, a giant $\gamma$-ray flare, SGR 1806-20,  has been observed.
The total flare
energy was estimated to be  as much as $2 \times 10^{46}$ erg.  This is $\sim 10^2$
times higher than the two previously observed  giant flares
\cite{Mazets1979,hurley1999}, and is believed to have been released during a reconfiguration
of magnetic fields of the neutron star. In Ref. \cite{cheng1996},
the similarities between SGR events and star-quakes relating to
the sudden change of pulsar periods were discussed, and the possibility
that SGRs may be powered by starquakes was suggested. These
phenomena might be explained by sudden changed in the magnetic pressure.

Although many uncertainties remain regarding  star-quakes and the
strength of magnetic fields in the interior of  neutron stars, it has been suggested
 \cite{su2001}  that neutron star matter
might be unstable in the presence of strong  magnetic fields because of the onset of discrete
Landau  level energy quantization.  This can  cause rapid changes in the pressure response to changes in density.  This instability could be
one source for star-quakes. Thus, one may conjecture the following
unifying scenario from star-quakes to pulsar glitches. If a
strong magnetic field can cause star-quakes, it may crack the
surface of the star.  Magnetic field energy may then be released
through the cracks.
The released magnetic energy and associated reconnection  may be observed as a SGR or a  giant flare. In
addition, the release of the magnetic field energy density may affect the equation
of state (EOS) of the neutron star.  This,  in turn, could  give rise to a  change of the moment of inertia, and thus, the
neutron-star spin period .

In Ref. \cite{su2001} properties of magnetic stars were studied, but only in the context of a cold free $n,p,e$ Fermi gas.  It could not be determined in that
work whether the features of interest would persist with a realistic nuclear equation of state. It is useful, therefore,
to reconsider the structure and dynamics of magnetic neutron stars   in the context of a realistic nuclear equation of state.
Therefore, in this paper we calculate the populations of particles, the
instability in the adiabatic index $\Gamma$, the EOS, and the moment of
inertia for various magnetic field strengths by using the techniques of quantum hadrodynamics
(QHD).  We then  discuss the possibility that  star-quakes and SGR flares might  be explained by the rapid change
of the adiabatic index $\Gamma$ with density due to the population of Landau levels.
We find that, similar to the Fermi gas approximation  \cite{su2001}, the magnetic QHD model also shows
an instability, {\it i.e.} a sudden change
of the   adiabatic index as the density increases.  As in the free nucleon gas this  is attributable to  the discrete  excitation  of Landau levels as the density increases, but also
to the appearance of hadronic species at high density.
These changes in the adiabatic index  lead to a sudden change in the pressure response of the star. Therefore,
star-quakes and an associated release of magnetic field energy may   take place.
Moreover, when we assume that  magnetic fields as large as
$~10^{18}$G  exist inside the star, we find that a release of  magnetic field energy could   decrease of the  moment of inertia leading to an increase in the the spin rate of the star.

In section II, we briefly introduce our theoretical framework for magnetized dense matter based upon the  QHD approach.
A method for calculating the possible change in the moment of inertia by the abrupt variations of the adiabatic index is also presented. Discussions of star-quakes and  the neutron-star
spin  are presented along  with numerical results in section III. A summary and conclusions are given in section IV.

\section{Theory }
\subsection{Relativistic mean field Lagrangian with strong magnetic fields}
The Lagrangian of the QHD model for dense nuclear matter in  the presence of  a magnetic
field can derived by introducing of a vector potential $A^{\mu}$.
The resulting Lagrangian can be written  in terms of the baryon octet,
leptons, and five meson fields as follows
\be \lag &=& \sum_b \bar \psi_b \Big [ i \gamma_\mu \partial^\mu
 - q_b \gamma_\mu A^\mu - M_b^*(\sigma, \sigma^*)  \nonumber \\
 &&- g_{\omega b} \gamma_\mu \omega^\mu
 - g_{\phi b} \gamma_\mu \phi^\mu  \nonumber \\
&& - g_{\rho b} \gamma_\mu \vec \tau \cdot \rho^\mu
 - \frac 12 \kappa_b \sigma_{\mu \nu} F^{\mu \nu}
\Big ] \psi_b  \nonumber \\
 &&+ \sum_l \bar \psi_l \left [  i \gamma_\mu \partial^\mu
 - q_l \gamma_\mu A^\mu - m_l \right ] \psi_l  \nonumber \\
&& + \frac 12 \partial_\mu \sigma \partial^\mu \sigma - \frac 12 m_\sigma^2 \sigma^2 - U(\sigma)  \nonumber \\
&&+ \frac 12 \partial_\mu \sigma^* \partial^\mu \sigma^* - \frac 12 m_{\sigma^*}^2 {\sigma^*}^2
- \frac 14 W_{\mu \nu}W^{\mu \nu} \nonumber \\
&& + \frac 12 m_\omega^2 w_\mu w^\mu
- \frac 14 \Phi_{\mu \nu} \Phi^{\mu \nu} + \frac 12 m_\phi^2 \phi_\mu \phi^\mu  \nonumber \\
&&- \frac 14 R_{i \mu \nu} R_i^{\mu \nu} + \frac 12 m_\rho^2 \rho_\mu \rho^\mu
- \frac 14 F_{\mu \nu} F^{\mu \nu},
\ee
where the indices $b$ and $l$ denote the baryon octet  and the leptons ($e^-$
and $\mu^-$), respectively. The effective baryon mass,
$M_b^*$, is simply given by $M_b^* = M_b - g_{\sigma b}\sigma -
g_{\sigma^* b} \sigma^*$, where $M_b$ is the free mass of a baryon
in vacuum and the $g_{\sigma b}$ are associated coupling constants. The $\sigma$, $\omega$ and $\rho$ meson fields describe
the nucleon-nucleon ($N-N$) and nucleon-hyperon
($N-Y$) interactions. The $Y-Y$ interaction is mediated by the $\sigma^*$ and $\phi$
meson fields. $U(\sigma)$ is the self interaction of the $\sigma$
field given by $U(\sigma) = \frac 13 g_2 \sigma^3 + \frac 14 g_3
\sigma^4$. $W_{\mu\nu}$, $R_{i \mu \nu}$, $\Phi_{\mu\nu}$, and
$F_{\mu\nu}$ denote  the  field tensors for the  $\omega$, $\rho$, $\phi$
and photon fields, respectively.

The anomalous magnetic moments (AMMs) of the baryons interact with the
external magnetic field via terms of the form of $\kappa_b \sigma_{\mu\nu}
F^{\mu\nu}$ where $\sigma_{\mu \nu} = \frac i2 [\gamma_\mu,
\gamma_\nu]$ and $\kappa_b$ is the strength of the AMM for  a baryon,
i.e. $\kappa_p = 1.7928 \mu_N$ for a proton with $\mu_N$ the
nuclear magneton.

In general, the AMMs could depend upon the matter density.
Therefore, one can take account of the medium effect through
density dependent AMMs which can be evaluated within the quark
meson coupling (QMC) model \cite{Ryu:2010zzb}. In this report, however,
we did not take account of the effect of density dependent AMMs because the calculation here is performed
within the QHD model.

Energy spectra for the  baryons and  leptons are given by
\be E_b^C &=& \sqrt{k_z^2 + \left ( \sqrt{{M_b^*}^2 + 2 \nu |q_b|
B} - s \kappa_b B \right )^2} \nonumber \\
&&+ g_{\omega b}\omega_0
+ g_{\phi b} \phi_0 + g_{\rho b} I_z^b \rho_{30} ~~, \nonumber \\
E_b^N &=& \sqrt{k_z^2 + \left ( \sqrt{{M_b^*}^2 + k_x^2 + k_y^2} - s \kappa_b B \right )^2} \nonumber \\
&&+ g_{\omega b}\omega_0 + g_{\phi b} \phi_0 + g_{\rho b} I_z^b \rho_{30} ~~, \nonumber \\
E_l &=& \sqrt{k_z^2 + m_l^2 + 2 \nu |q_l| B},
\ee
where $E_b^C$ and $E_b^N$ denote the energies of  charged
and  neutral baryons, respectively. The Landau quantization for
charged particles in a magnetic field is denoted as $\nu = n +
1/2 - sgn(q) s/2 = 0, 1, 2 \cdots~$, where the  sign of the electric charge
is denoted as $q$ $sgn(q)$ and $s=1(-1)$ is for spin up (down).

Chemical potentials for
the baryons and leptons are given by \be \mu_b &=&
E_f^b + g_{\omega b}\omega_0 + g_{\phi b} \phi_0 + g_{\rho b}
I_z^b \rho_{30}, \label{eq:che-b}
\\
\mu_l &=& \sqrt{k_f^2 + m_l^2 + 2 \nu |q_l| B}, \label{eq:che-l}
\ee
where $E_f^b$ is the baryon Fermi energy  and $k_f$ is the lepton
Fermi momentum. For charged particles, $E_f^b$ is
written as
\be {E_f^b}^2 = {k_f^b}^2 + (\sqrt{{m_b^*}^2 + 2 \nu |q_b| B} - s
\kappa_b B)^2,
\ee
where $k_f^b$ is the baryon Fermi momentum  and $\nu $ = 0 for
neutral baryons.

We apply  three constraints for calculating
the properties of a neutron stars:  1) baryon number conservation; 2) charge
neutrality; and 3) chemical equilibrium. The meson field equations are
solved along with the chemical potentials for  baryons and leptons subject to these
 three constraints. The total energy density is then given by
$\varepsilon_{tot} = \varepsilon_m + \varepsilon_f$, where the
energy density for the matter fields is given as
\be \varepsilon_m &=& \sum_b \varepsilon_b + \sum_l \varepsilon_l
+ \frac 12 m_\sigma^2 \sigma^2 + \frac 12 m_{\sigma^*}^2
{\sigma^*}^2 \nonumber \\
&& + \frac 12 m_\omega^2 \omega^2 + \frac 12 m_\phi^2
\phi^2 + \frac 12 m_\rho^2 \rho^2 + U(\sigma),
\ee
and the energy density due to the magnetic field is given by
$\varepsilon_f = B^2 / 2$. The total pressure is then
 \be P_{tot} = P_m + \frac 12 B^2, \ee where the pressure due to the matter
fields is obtained from $P_m = \sum_i \mu_i \rho_v^i -
\varepsilon_m$. The relation between the mass and radius for a static
and spherical symmetric neutron star is generated from a solution to
the Tolman-Oppenheimer-Volkoff (TOV) equations using  the equation
of state (EoS) described above.

\subsection{Slowly rotating neutron stars}
To calculate the moment of inertia for a
slowly rotating neutron star,we follow the methods  detailed in Refs.
\cite{glen,Fattoyev:2010tb} which are briefly summarized here. The metric of an axially symmetric
neutron star can be written in generalized rotating Schwarzschild  coordinates as
\be
 ds^2 &=& g_{\mu\nu}dx^{\mu}dx^{\nu} =
-e^{2\nu(r)}dt^2 + e^{2\lambda(r)} dr^2 +
  r^2 d\theta^2 \nonumber \\
 &&+ r^2 \sin^2\theta d\phi^2 -
 2\omega(r)r^2\sin^2\theta dt d\phi \;.
 \label{metric}
\ee

If the  neutron star is rotating uniformly with
a stellar frequency $\Omega$ far below
the Kepler frequency
\be
  \Omega \ll \Omega_{\rm max} \approx \sqrt{\frac{M}{R^3}}~,
 \label{Kepler}
\ee
the slow-motion approximation is valid and the moment of inertia can be written:
\be
  I \equiv \frac{J}{\Omega} = \frac{8\pi}{3}
 \int_{0}^{R} r^{4} e^{-\nu(r)}\frac{\bar{\omega}(r)}{\Omega}
 \frac{\Big ( \varepsilon(r)+P(r) \Big)}{\sqrt{1-2M(r)/r}} dr \;,
 \label{MomInertia}
\ee
where $J$ is the angular momentum, while $\nu(r)$ and $\bar{\omega}(r)$
are radially-dependent metric functions. $M(r)$, $\varepsilon(r)$,
and $P(r)$ are the mass of the star, energy density, and pressure,
respectively, derived  from a solution to the  TOV equation.

The metric functions $\nu(r)$ and $\lambda(r)$ in Eq.
(\ref{metric}) are unchanged from the values for a spherically
symmetric neutron star. Thus, $\lambda(r)$ is obtained from the
mass $M(r)$ by the usual Schwarzschild solution:
\be
  g_{11}(r)=e^{2\lambda(r)}=\Big(1-2M(r)/r\Big)^{-1} \;
 \label{metric1}
\ee
and $\nu(r)$ can be calculated from
\be
 \nu(r) &=& \frac{1}{2}\ln\left(1-\frac{2M}{R}\right) \nonumber \\
   && -\int_{r}^{R} \frac{\Big(M(x)+4\pi x^{3}P(x)\Big)}
              {x^{2}\Big(1-2M(x)/x\Big)}dx \;.
\label{metric2}
\ee

The metric function $\omega(r)$ denotes  a frequency corresponding to the dragging of local
inertial frames by the rotating star. The relative frequency
$\bar{\omega}(r)\!\equiv\!\Omega\!-\!\omega(r)$ appearing in
Eq.~(\ref{MomInertia}) represents the angular velocity of the
fluid as measured in a local inertial reference frame. In
particular, the dimensionless relative frequency
$\widetilde{\omega}(r)\!\equiv\!\bar{\omega}(r)/\Omega$ satisfies
the following second-order differential equation
\be
 \frac{d}{dr}\left(r^{4}j(r)\frac{d\widetilde{\omega}(r)}{dr}\right)
 +4r^{3}\frac{dj(r)}{dr}\widetilde{\omega}(r) = 0\;,
 \label{OmegaBar}
\ee
where
\be
  j(r)=e^{-\nu(r)-\lambda(r)} =
    e^{-\nu(r)}\sqrt{1-2M(r)/r}  & \text{if $r \le R\;,$} \nonumber \\
    1 &\text{if $r > R\;.$} \nonumber \\
\ee
Note that $\widetilde{\omega}(r)$ is subject to the following two
boundary conditions:
\be
  \widetilde{\omega}'(0) &=& 0 \;,
  \label{BC1}\\
  \widetilde{\omega}(R) &+& \frac{R}{3}\,\widetilde{\omega}'(R)=1~.
  \label{BC2}
\ee
In  numerical calculations, one can integrate Eq.~(\ref{OmegaBar}) from
the center of the star to its surface for various  central frequencies.
When the boundary conditions at the surface, Eq.~(\ref{BC2}) are
not  satisfied for an arbitrary choice of $\widetilde{\omega}_{c}$,
one must rescale the function and its derivative by an appropriate
constant to correct for the mismatch. After solving both $\widetilde{\omega}(r)$ and $I$,
one can check the consistency of
the results  by testing the accuracy to which the  following condition is satisfied:
\be
  \widetilde{\omega}'(R) = \frac{6I}{R^{4}} \;.
  \label{omegaR}
\ee

\section{Results and discussion}

We use the parameter set given in Ref. \citep{RHHK} for the coupling
constants, $g_{\sigma N}$, $g_{\omega N}$ and $g_{\rho N}$, where
$N$ denotes the nucleon. The coupling constants for hyperons in
a nuclear medium, $g_{\omega Y}$ and $g_{\sigma Y}$, are determined
by the quark counting rule and the relevant hyperon potentials at the
saturation density.  The strength of these potentials is fixed at $U_\Lambda = -30$
MeV, $U_\Sigma = 30$ MeV and $U_\Xi = -15$ MeV. Since the magnetic
fields may also depend upon the density, we utilize  the form for the density-dependent
magnetic fields  suggested in  Refs.
\citep{band1997,panda2009,Ryu:2010zzb}
\be B \left ( \rho / \rho_0 \right ) = B^{surf} + B_0 \left [ 1 -
\exp \{-\beta \left ( \rho / \rho_0 \right )^\gamma \} \right ],
\label{eq-B} \ee
where $B^{surf}$ is the magnetic field at the neutron-star surface,
 taken from observations to be  $\sim 10^{15}$ G, and $B_0$
represents the magnetic field saturation strength in the high density
region.

In this work, we use $\beta = 0.02$ and $\gamma = 3$. Since the
magnetic field is usually written in units of the  critical field for the
electron $B_e^c = m_e^2 / e = 4.414 \times 10^{13}$ G, the $B$ and the $B_0$
in Eq. (\ref{eq-B}) can be expressed  as dimensionless quantities $B^* = B / B_e^c$ and $B_0^*
= B_0 /B_e^c$. Based upon this magnetic field representation, we investigate the
structure of neutron stars as a function of magnetic
field strength both with and without hyperonic matter.

\subsection{Particle populations, adiabatic index and equation of state}

\begin{figure}
\centering
\includegraphics[width=7.5cm]{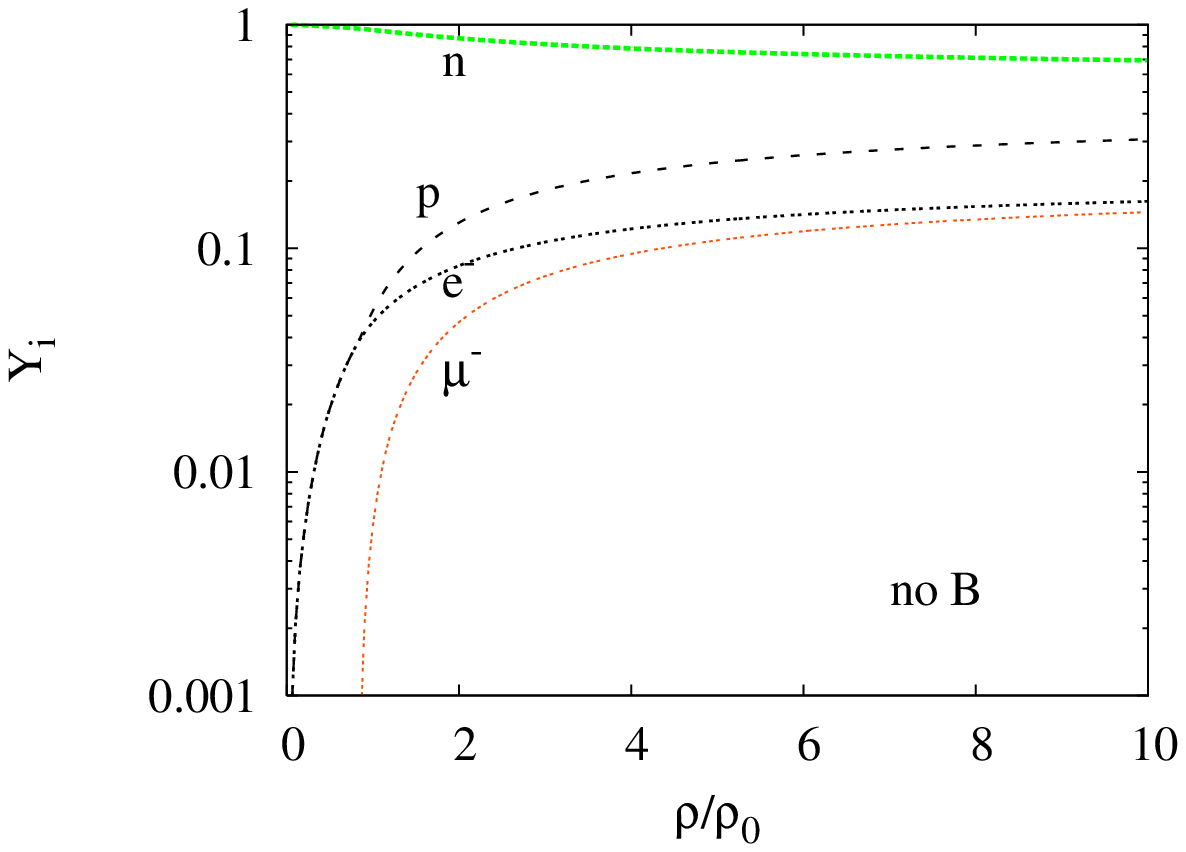}
\includegraphics[width=7.5cm]{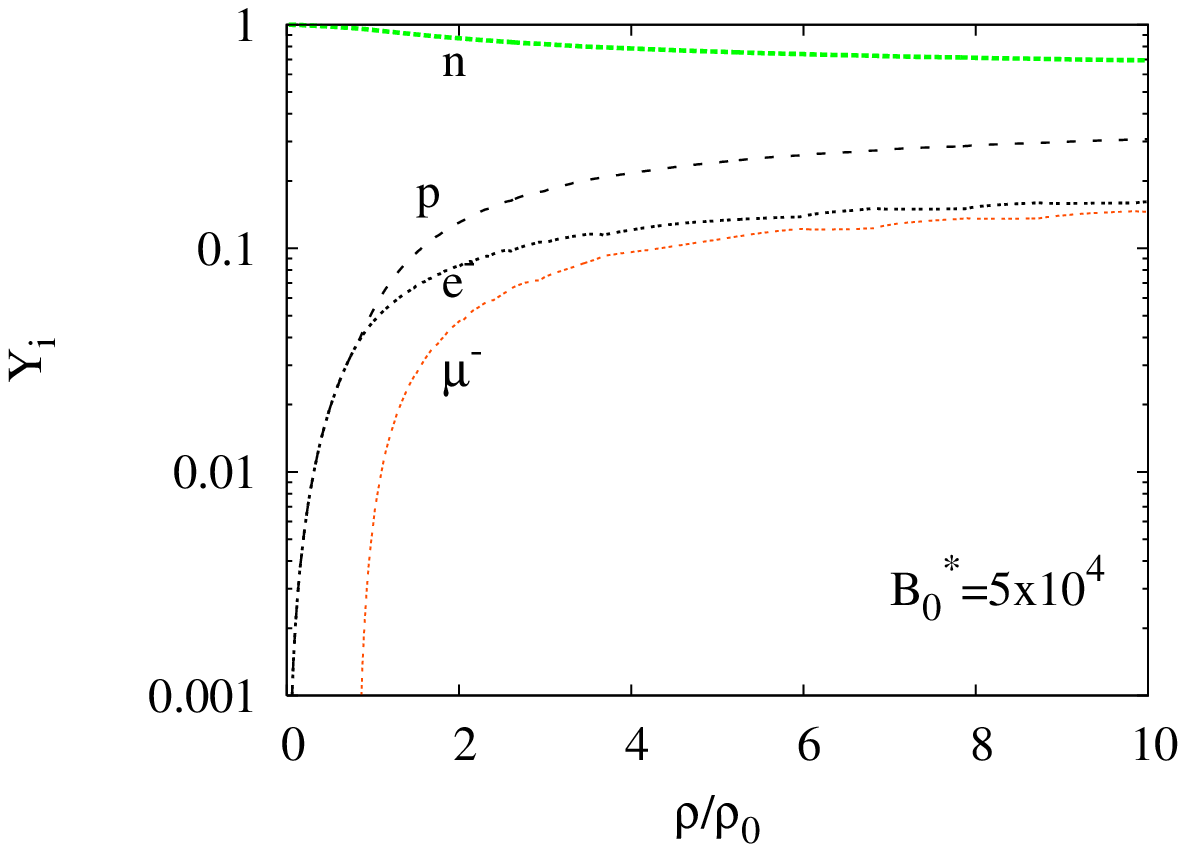}  \\
\includegraphics[width=7.5cm]{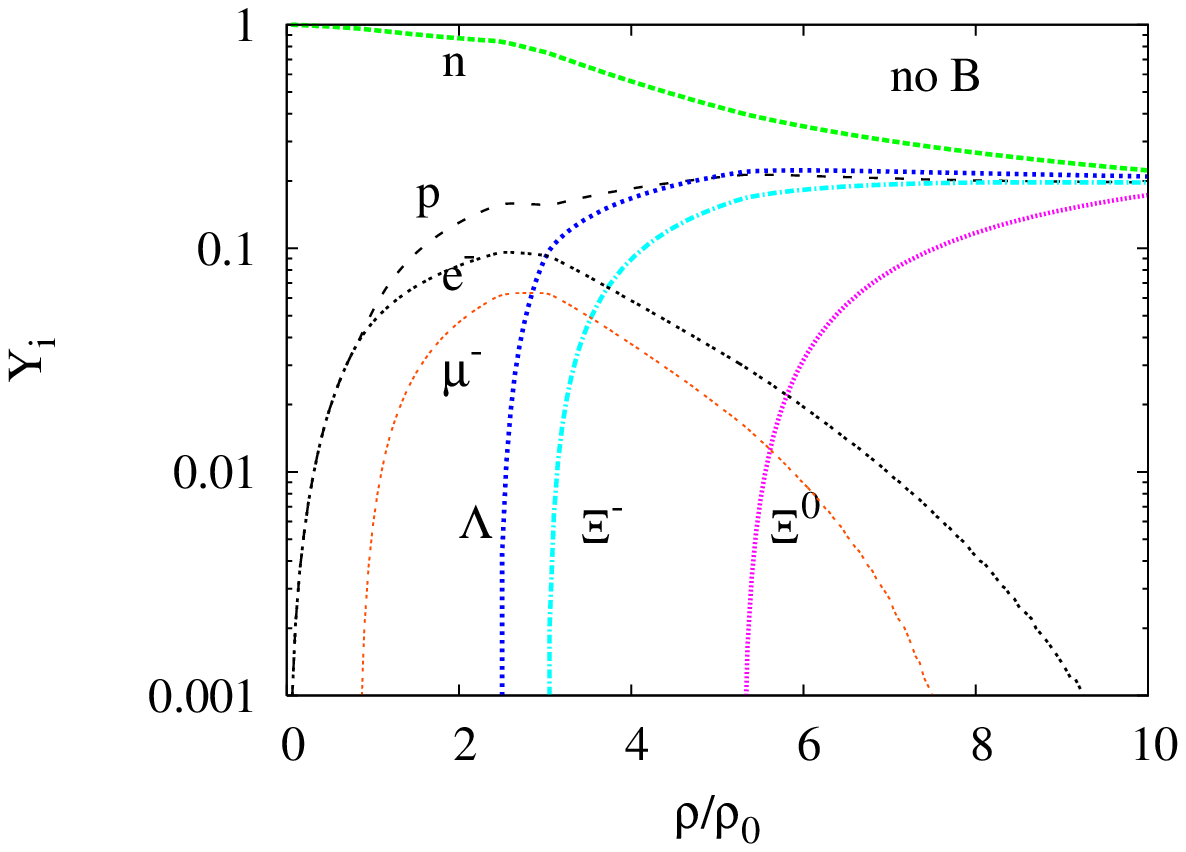}
\includegraphics[width=7.5cm]{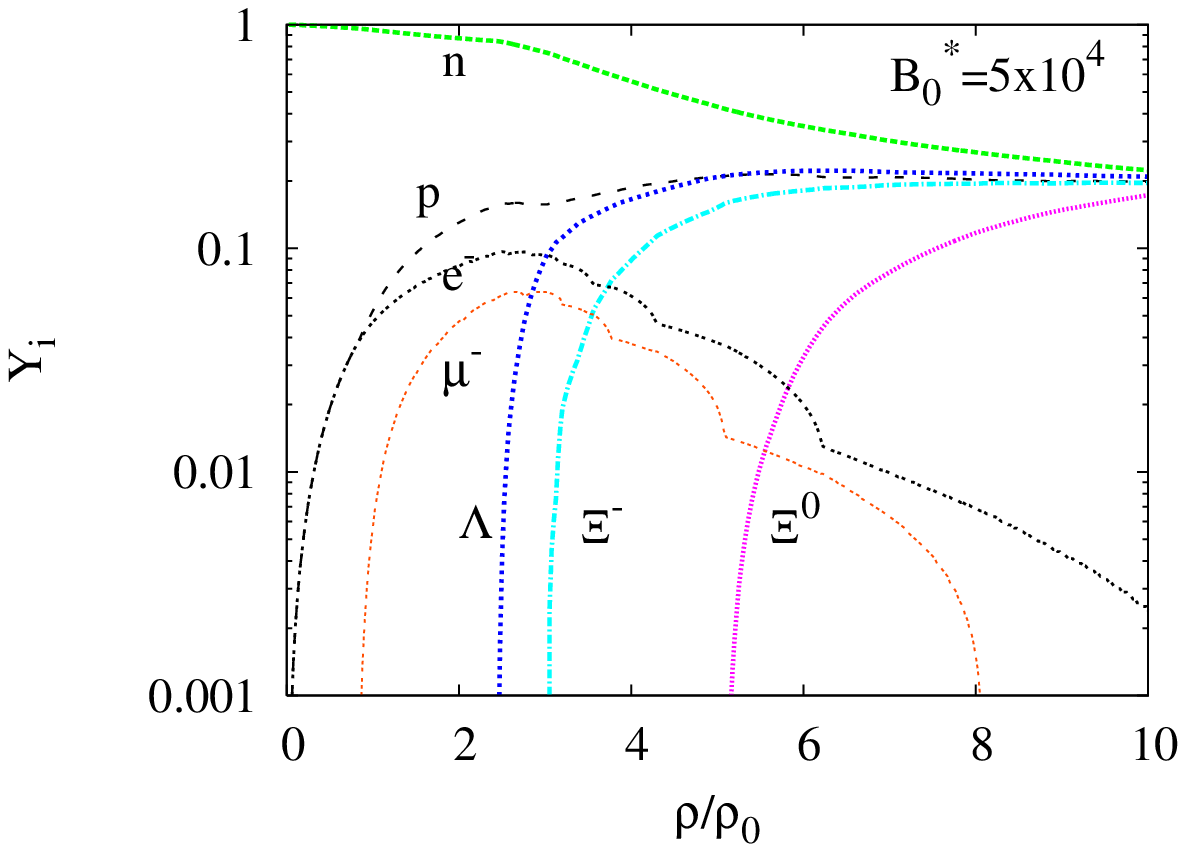}
\caption{Populations of particles in a neutron star. The upper two panels
are for the nucleonic phase and the lower two panels are for the hyperonic phase
with and without a magnetic field, respectively, where we used $B_0^* = 5 \times
10^4$ as explained in the  text.  Note, this field strength is a bit different value from the value used in our previous
paper \cite{Ryu:2010zzb}.} \label{fig:popul}
\end{figure}

The populations of particles as a function of density in our model are shown in Fig. \ref{fig:popul},
for four cases.  Shown are particle fractions $Y_i$ for  both  pure nucleonic and hyperonic equations of state, and both with and without
magnetic fields. Since the details about the
effects of magnetic fields in neutron stars have been  discussed in our
previous paper \cite{Ryu:2010zzb}, we briefly explain the magnetic field effects on populations of particles. In Fig.
\ref{fig:popul}, one can notice the effects of a magnetic field
from the difference between the  (no $B$) and   ($B_0^* = 5 \times 10^4$) figures  for both nucleonic (upper)
and hyperonic phases (lower).

In general, charged particles are strongly affected by the EM
interaction term ($eB$), which gives rise to an increase in
the electron fraction. Therefore the condition of charge neutrality and the EM
interaction enhances the proton fraction.  This  suppresses
hyperons by  baryon number conservation. This phenomenon explicitly appears if we utilize $B_0^* = 10^5$ as in our previous paper \cite{Ryu:2010zzb}. But its effect becomes indiscernible in the present case with a weaker field strength, $B_0^* = 5 \times 10^4$.

The most prominent feature caused by the strong magnetic field in Fig. 1 is the kink pattern in the populations of electrons and muons, which is clearly shown for the hyperonic phase. Of course, that pattern is caused by the change of Landau levels owing to the existence of a magnetic field.

\begin{figure}
\centering
\includegraphics[width=7.8cm]{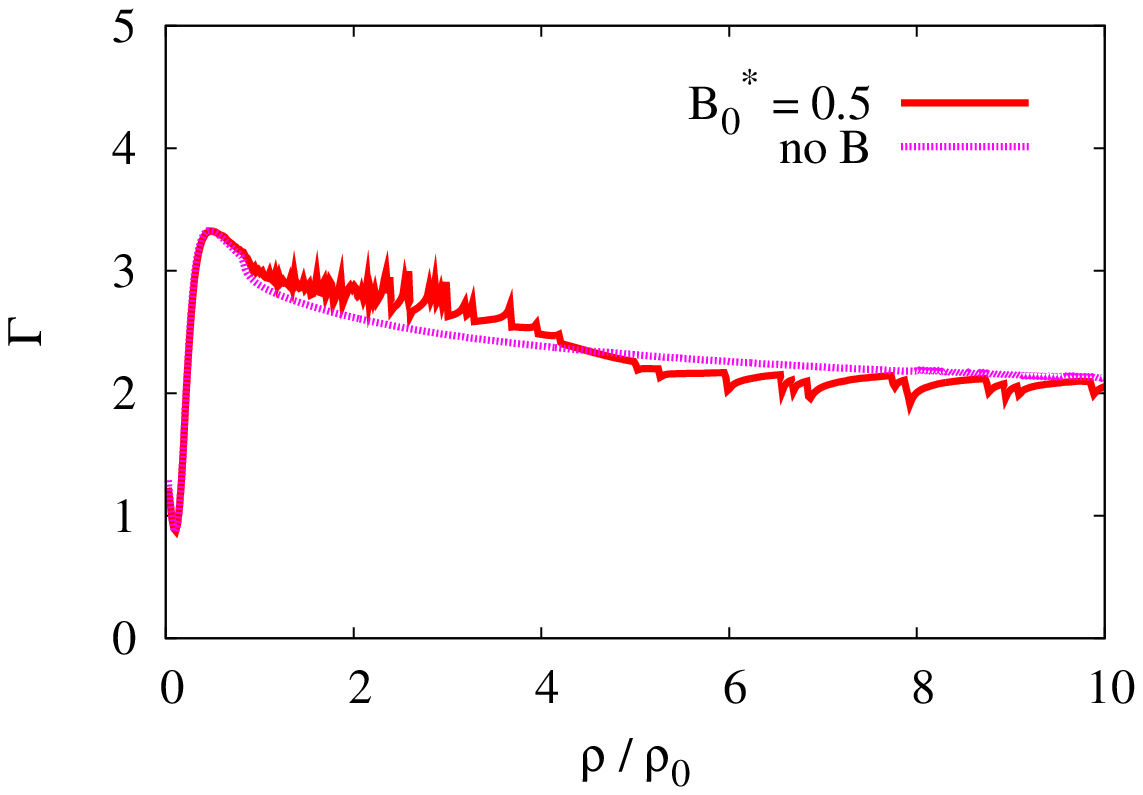}
\includegraphics[width=7.8cm]{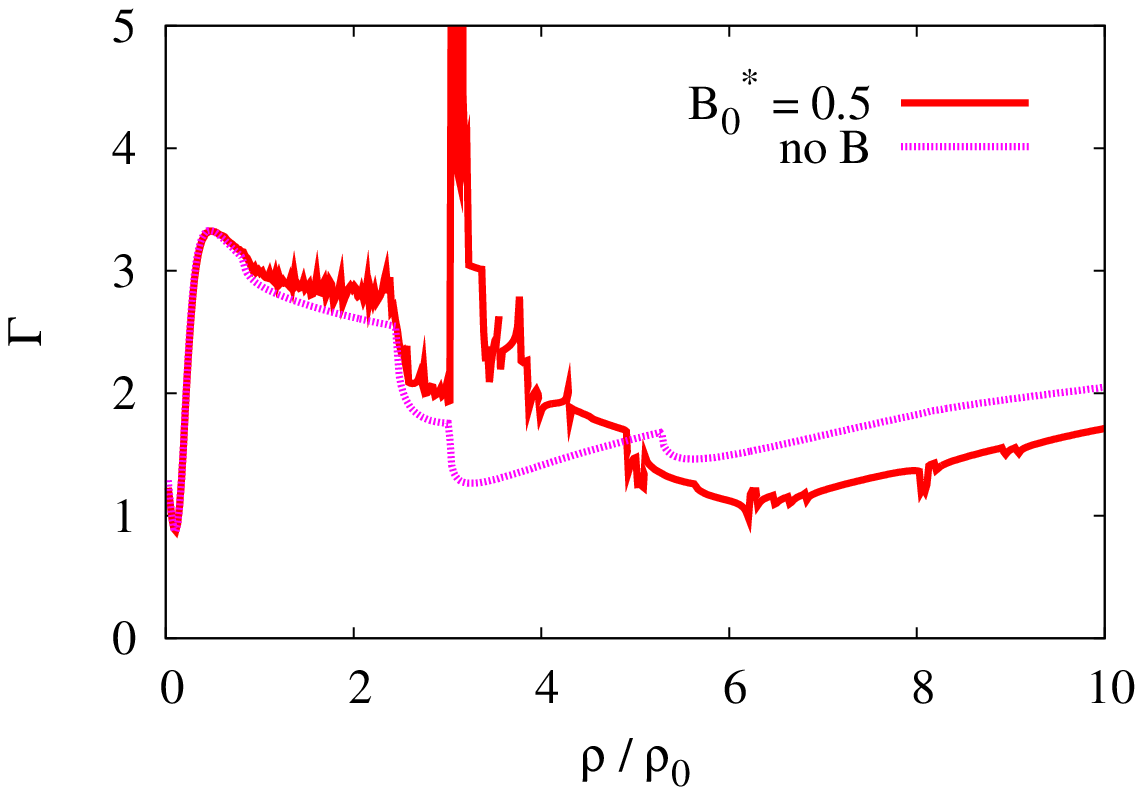}
\caption{Adiabatic index for the nucleonic  (upper) and hyperonic
(lower) equations of state. The adiabatic index is defined as $\Gamma \equiv
\frac{\partial \ln P}{\partial \ln \rho} = (1 + \frac \varepsilon P )
\frac{dP}{d\varepsilon}$. Here $B_0^*$ is given in  units of
$10^4$.} \label{fig:adia}
\end{figure}
\begin{figure}
\centering
\includegraphics[width=7.8cm]{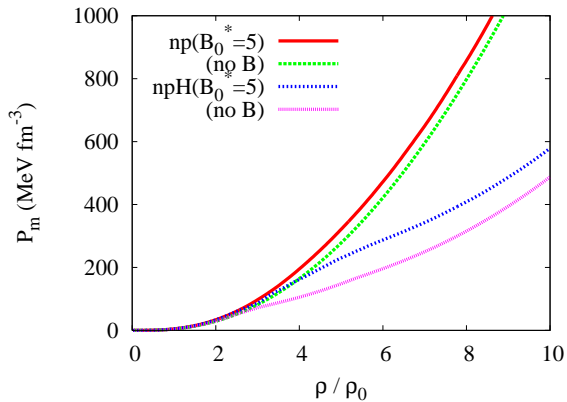}
\caption{Equation of state pressure as a function of density in units of the nuclear saturation density. }
\label{fig:eos}
\end{figure}

In Fig. \ref{fig:adia}, we show the adiabatic index $\Gamma \equiv
\frac{\partial \ln P}{\partial \ln \rho} = (1 + \frac \varepsilon P )
\frac{dP}{d\varepsilon} $ for both the nucleonic (upper) and the hyperonic (lower)
equation of state.  This is of particular interest  because the adiabatic index may be used as a critical
factor for understanding the radial stability of a star \cite{Shapiro83,Shapiro91}.

In the nucleonic phase, one can see multiple kinks similar to the
results shown in Ref. \cite{su2001} based upon a cold Fermi gas model. This
oscillatory pattern takes place as charged particles begin to
rapidly fill the next unoccupied Landau level when the density increases.
In the hyperonic phase, one can see a sudden large increase in the adiabatic index
at  $\approx 3\rho_0$.  This is caused by the production of $\Xi^-$
hyperons around that  density as shown in Fig. 1.

In the high density region beyond $3\rho_0$, similar kink patterns appear in both the nucleonic and
hyperonic phases. At high density, however,  the kinks are relatively
smaller  than the kinks in the lower density region.  This is  because
the high density increases the occupied levels but does not cause
a rapid shift in the Landau levels. This  means that there should exist a critical density for such discontinuous pressure response, at least in the hyperonic phase.

Similarly to the Fermi gas model \cite{su2001}, the magnetic fields in the QHD model also give rise to  sudden jumps from the occupied Landau levels to the unoccupied levels.  This leads to  kinks in the adiabatic index because of the unstable
structure in the magnetic pressure. These oscillatory kinks can be conjectured to cause
star-quakes. Specifically, in the hyperonic phase, the instability caused by the magnetic field becomes significantly larger than that in the nucleonic phase.

\begin{figure}
\centering
\includegraphics[width=7.8cm]{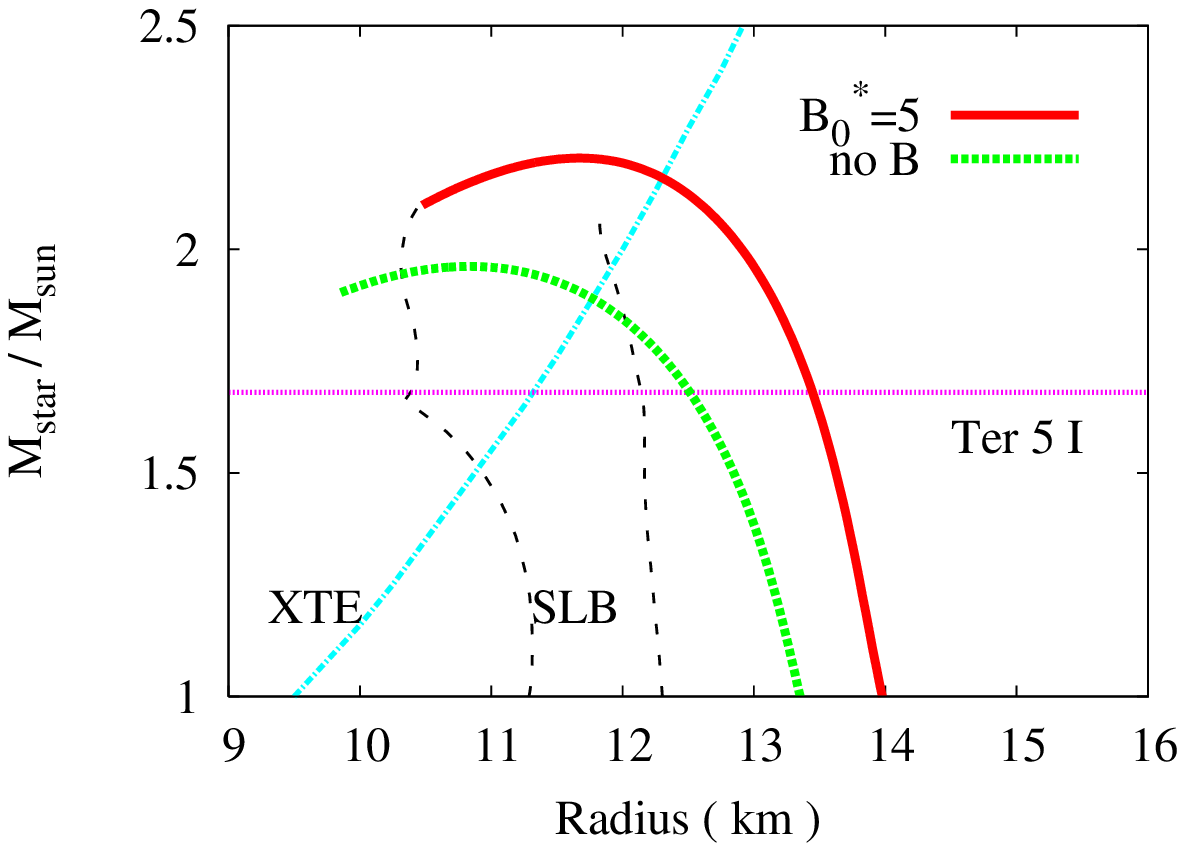}
\includegraphics[width=7.8cm]{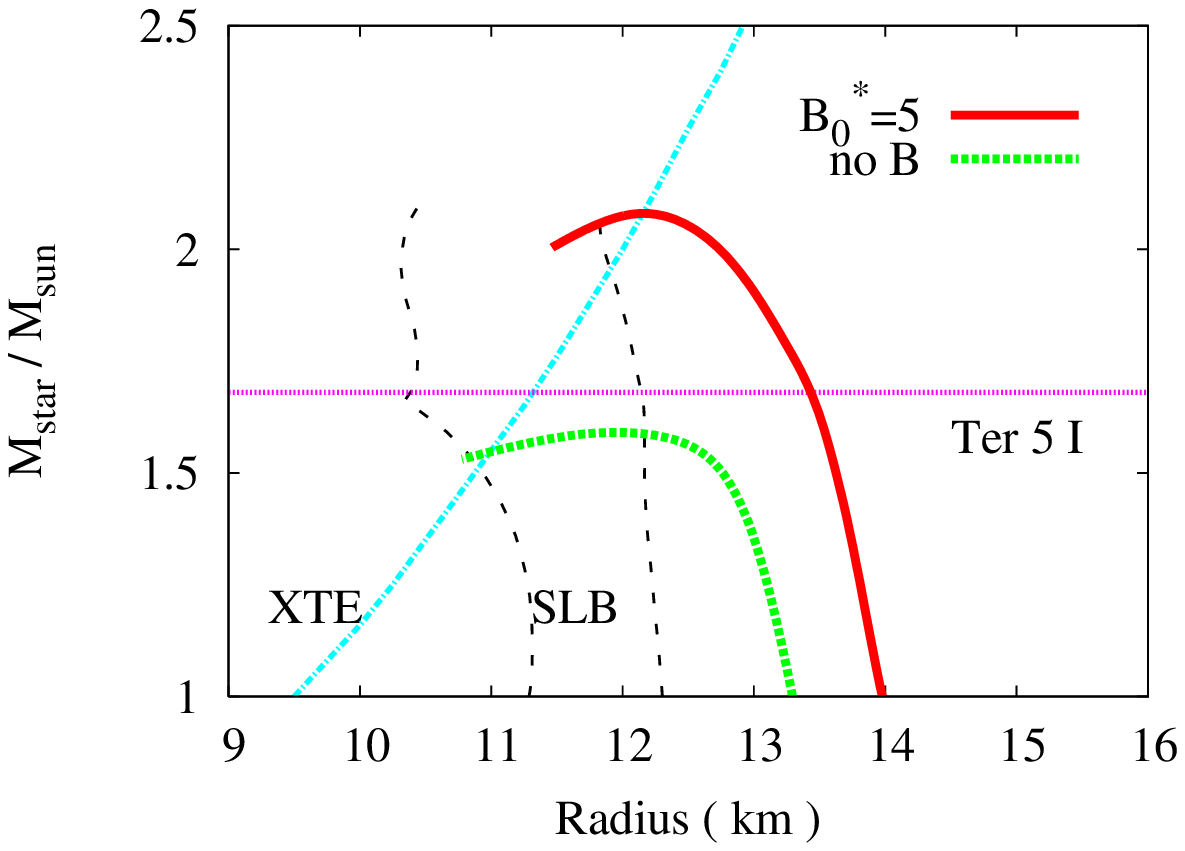}
\caption{Mass-radius relations for nucleonic (upper) and
hyperonic (lower) equations of state. Three constraints are shown in the figure.
XTE is a short expression of XTE J1739-285 \cite{kaaret}, Ter 5 I
is Pulsar I of the globular cluster Terzan 5($M_{star} = 1.68
M_\odot$) \citep{ransom}, and SLB is taken from the data
($2\sigma$ limits) on $r_{ph} \gg R$ in Ref. \cite{Steiner:2010fz}
where $r_{ph}$ is the photospheric radius. $B_0^*$ values are
given in units of $10^4$. Detailed explanations are given in
text.} \label{fig:mr}
\end{figure}

In Fig. \ref{fig:eos}, the EOSs are shown for both nucleonic and
hyperonic matter. Magnetic fields can make the EOS stiff in both
phases. In particular, since the magnetic fields suppress
hyperons, the effect of stiffness owing to the magnetic field in
the hyperonic phase is far larger than in the nucleonic phase. Thus, the stiff EOS gives a large mass and radius for a neutron star.
That is shown in Fig. \ref{fig:mr}.

\subsection{Mass-radius relation and the moment of inertia}
%
\begin{figure}
\centering
\includegraphics[width=7.8cm]{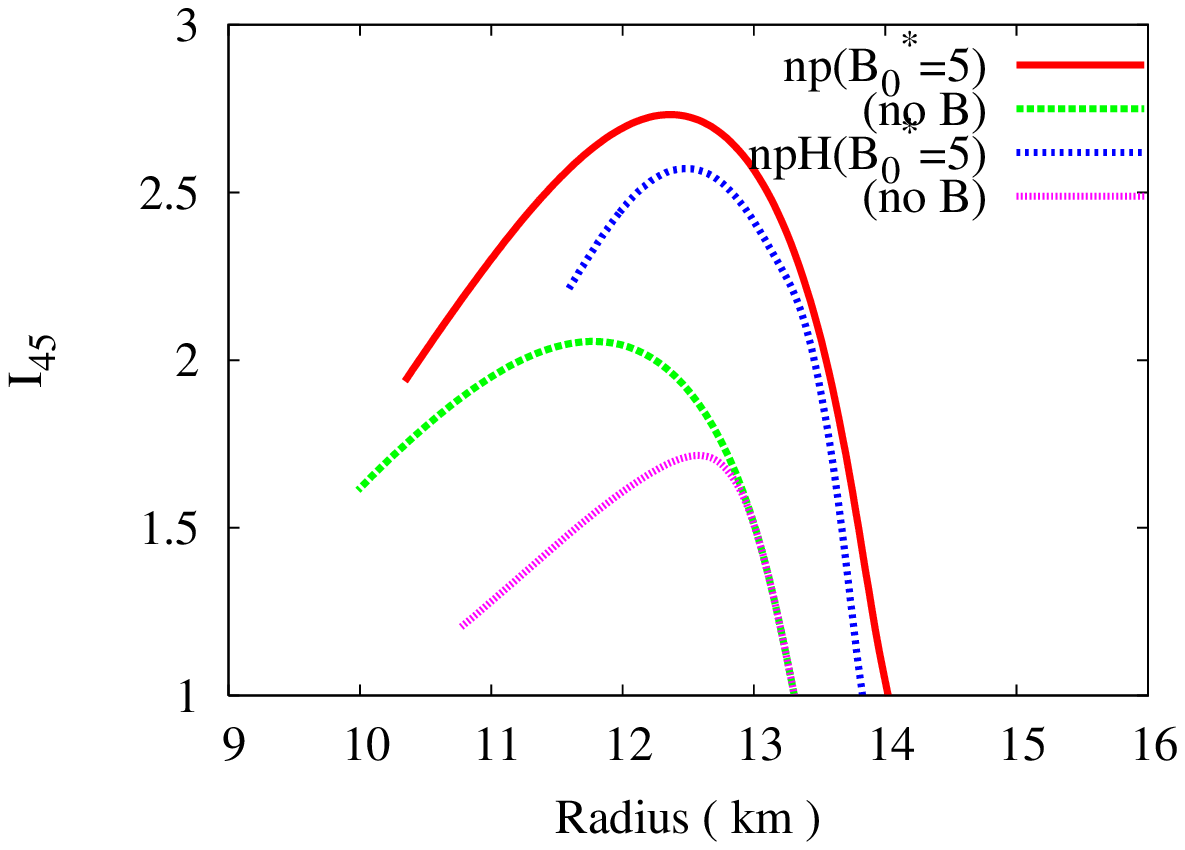}
\includegraphics[width=7.8cm]{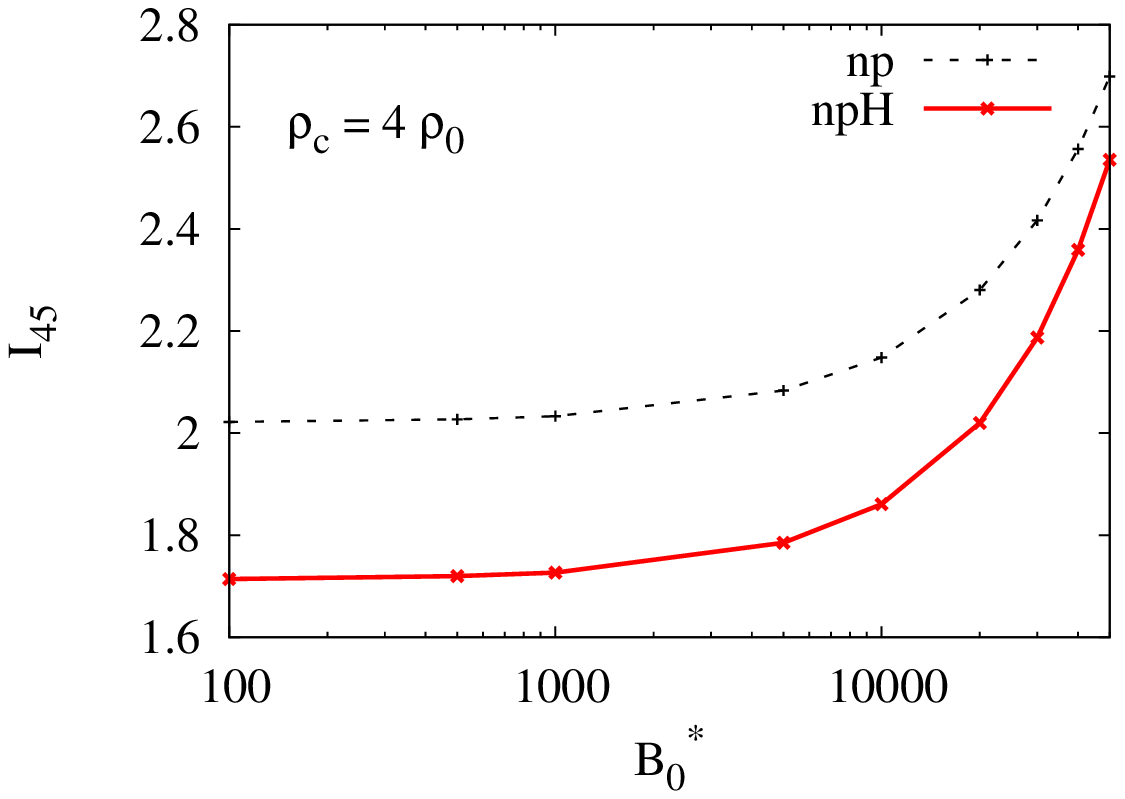}
\caption{The moment of inertia ($I_{45} = 10^{45}$ g $\cdot$
cm$^2$) for nucleonic and hyperonic equations of state  with and without
magnetic fields as a function of  of radius (upper) and magnetic field strength
(lower) figure.} \label{fig:imom}
\end{figure}

The mass-radius relations for both the nucleonic   and
the hyperonic   phases are compared with  observational limits   in Fig. 4.
In this context, it is particularly noteworthy that larger maximum masses and radii
are possible when a strong interior magnetic field is present.
Neutron
stars and heavy ion collision data provide valuable constraints on
the EOS for dense matter \cite{klahn2006}. There are indeed  recent  reports of
evidence for higher maximum masses for neutron stars. For instance, $M = 2.0 \pm 0.1
M_\odot$ for 4U 1636-536 was reported in Ref. \citep{lava2008}.
Very recently authors in Ref. \cite{Steiner:2010fz} investigated seven
neutron stars, six binaries and an isolated neutron star (RX
J1865-3754) and showed that $M \approx 1.9-2.3 M_\odot$ and $R \approx 11
- 13$ km. That is labeled  as the SLB data in Fig. 4, which also shows the  $2\sigma$ lower and upper limits.

Pulsar I of the globular cluster Terzan 5 (Ter
5 I) shows a lower mass limit $M_{N.Star} \ge 1.68 M_\odot$ at the 95
\% confidence level \citep{ransom}. Another constraint deduced
independently of given models is obtained from XTE J1739-285
\citep{kaaret}.  That work provides  a constrained curve for the ratio
between mass and radius. Although we do not indicate it here, another
recent paper \cite{demorest2010} has also reported a mass of $M \approx 1.97 M_\odot$
for PSR J1614-2230 from a detection of the  Shapiro time delay.

For the case of  a  nucleonic EOS, the mass and radius relations satisfy
all constraints.  The  hyperonic phase, however,  shows some discrepancies.
Although a hyperonic phase cannot satisfy all of the constraints,
magnetized hyperonic neutron stars can explain
large neutron-star masses within the constraints from the observational data.

In Fig.~\ref{fig:imom}, we show the moment of inertia in
units of $I_{45} = 10^{45}$ g cm$^2$ in order to investigate the spin up of the star. One can easily
notice the large difference in the moment of inertia between
stars with and without magnetic fields. This  implies that reducing the
magnetic field strength  could cause a decrease in the moment of
inertia. If there is a mechanism for this reduction, one might expect an decrease in the spin period by  angular momentum
conservation.

For illustration, we plot  $I_{45}$ as a function of $B_0^*$ on
 Fig. \ref{fig:imom} for  models with  a central density of $\rho
= 4 \rho_0$. For a can field strength around $B_0^* = 10^4$,  relatively small changes of
the magnetic field give rise to a significant change in the moment of inertia
leading to a  spin change of  the neutron stars.



To examine the effect of SGR flares on the moment of inertia, consider the released magnetic energy  is $E_{flare} \sim 2 \times 10^{46}$ erg  for the largest observed magnetar flare yet observed  \cite{palmer2005}.
If we assume that this magnetic energy released from a region below the surface of thickness of thickness $\Delta r$, than the released energy is related to the change in the magnetic energy density $(\Delta B)^2/2)$ times a volume factor,  $E_{flare} = (4 \pi R^2 \Delta r) \times (\Delta B)^2 /2$
where $R$ is the radius of the emission region taken to be  $\sim 10$ km. An emission $\sim 10^{46}$ erg  corresponds to a change in the magnetic field of only  $\Delta B \sim  10^{16}$ G from a thickness of $\Delta r= 1$ m ( or $\sim 10^{15}$ G for $\Delta r =$  100 m ).  From Fig.~\ref{fig:imom} it is apparent that a change of $\sim 10^{16}$ G corresponds to  about a 1\% decrease in the moment of inertia and the rotation period.  For a typical flare associated with AXPs, however, the energy released is only $\sim 10^{40}$ erg \cite{Mereg09}, corresponding to $\Delta B \sim 10^{13}$ G, implying a decrease of the rotation period by a factor  of $\sim 10^{-5}$.  Interestingly, this is comparable  to the glitches observed \cite{Dib08} in some AXPs in association with a radiative event.  Thus, an association of AXP emission and glitches is possible in this mechanism.

\section{Summary}
We have investigated neutron stars which contain  strong interior magnetic fields.
The populations of particles, the adiabatic index, the EOS, the mass-radius
relation, and the moment of inertia have been calculated by using the QHD
Lagrangian  modified to include strong magnetic fields.  We  discuss the effects  of a strong interior magnetic field  on the
physical quantities characterizing neutron stars or magnetars.

Possible changes in the  mass-radius
relation for neutron stars with strong magnetic fields are
compared with the constraints deduced from observational data.
The mass-radius relations for both nucleonic and hyperonic phases
with strong magnetic fields satisfy the maximum mass suggested  by Ter 5 I and
XTE J1739-285, but the constraint in Ref. \cite{Steiner:2010fz}
turns out to be inconsistent with the results for the  hyperonic phase unless a strong interior magnetic field is present.

The possibility of star-quakes,  magnetar flares and changes in the spin rate of AXPs associated with flares has  also been addressed
by determining  the adiabatic index and the moment of inertia
with a magnetic  QHD equation of state. Our results suggest  that   changes in the pressure response  of the
star with changes in density could lead to star quakes and magnetic flare energy release consistent with magnetar flares if the
interior magnetic field strength is large enough.  The implied change in the moment of inertia associated with
magnetic flares is also consistent with glitches  in the AXP period occasionally associated with radiative events.

\acknowledgments This work was supported by the National Research Foundation of Korea (Grant No. 2011-0077273). Work at NAOJ was supported by Grants-in-Aid for
 Scientific Research of the JSPS (20244035), and for Scientific Research on
 Innovative Area of MEXT (20105004). Work at the University of Notre Dame was
 supported by the U.S. Department of Energy under Nuclear Theory Grant
 DE-FG02-95-ER40934.

\end{document}